\documentclass[aps,onecolumn,floatfix,nofootinbib]{revtex4}

\usepackage{graphicx,epsfig,wrapfig,color}
\usepackage{amsmath,amssymb,dsfont,hyperref}
\usepackage[normalem]{ulem}
\usepackage{xcolor,wasysym,tikzsymbols}
\usepackage{color}

\newcommand{\be}{\begin{equation}}
\newcommand{\ee}{\end{equation}}
\newcommand{\ba}{\begin{eqnarray}}
\newcommand{\ea}{\end{eqnarray}}

\newcommand{\la}{\langle}
\newcommand{\ra}{\rangle}
\newcommand{\rhofl}{\rho_{\mbox{\tiny fluid}}}
\newcommand{\pfl}{p_{\mbox{\tiny fluid}}}
\newcommand{\mfl}{m_{\mbox{\tiny fluid}}}

\begin{document}
\newcommand*{\ZZ}{
    $^1$ Department of Physics, Illinois Institute of Technology, Chicago, Illinois 60616-3793, USA\\
    $^2$ Department of Physics, University of Connecticut, Storrs, CT 06269-3046, USA\\
    $^3$ Department of Physics, Yale University, New Haven, CT 06520, USA
    }\affiliation{\ZZ}
\title{The energy-momentum tensor in a classical model of the electron}
\author{Grace Gardella$^{1,2}$, Mira Varma$^{3,2}$, Peter Schweitzer$^2$}\affiliation{\ZZ}
\date{March 2026}
\begin{abstract}
We show that the leading non-analytic terms in the small-$t$ expansion of the energy momentum 
tensor (EMT) form factors of an electrically charged particle in QED can be correctly derived 
in a classical model of the electron by Bia\l ynicki-Birula. 
Based on the lucidity of the employed exactly solvable model, we comment also on the recently 
proposed concept of a regularized proton~$D$-term.
\end{abstract}
\maketitle

\section{Introduction}

Form factors of the energy-momentum tensor (EMT) have attracted considerable attention in the 
recent past. The EMT matrix elements give access to fundamental particle properties including the 
mass \cite{Ji:1995sv}, spin \cite{Ji:1996ek} and $D$-term \cite{Polyakov:1999gs} with significant
potential for new insights in hadron structure
\cite{Ji:1998pc,Radyushkin:2000uy,Goeke:2001tz,Diehl:2003ny,Belitsky:2005qn,Boffi:2007yc}.
In theoretical studies of hadrons, the focus is 
naturally on the short-range strong forces. Electromagnetic effects, often deemed to play no major 
role for hadron structure, are rarely considered in such studies. In systems governed by short-range
forces one finds a negative EMT form factor $D(t)$ which has a regular limit $D=\lim_{t\to0}D(t)$ 
known as the $D$-term.\footnote{The form factor $D(t)$ \cite{Polyakov:1999gs} 
    has attracted a lot of interest also due to its appealing interpretation in 
    terms of mechanical properties \cite{Polyakov:2002yz}, see the reviews 
    \cite{Polyakov:2018zvc,Burkert:2023wzr}. The interpretation was elaborated 
    in \cite{Lorce:2018egm,Freese:2021czn} and encountered also some criticism in 
    literature \cite{Lorce:2025oot,Ji:2025qax,Miller:2025zte}. 
    In this work we use the interpretation in a ``reversed direction'' as a practical
    tool for our calculations and demonstrate that we obtain correct results. This is
    not intended to resolve any controversy, but it speaks for itself: when applied
    responsibly within its limitation, the 3D interpretation yields correct results
    for form factors. This is how Hofstadter used the 3D interpretation in his 
    seminal works \cite{Hofstadter:1956qs}.}

The presence of electromagnetic or other long-range forces affects the EMT properties 
strongly. In particular, the form factor $D(t)$ of a charged particle develops a $1/\sqrt{-t}$ 
singularity and the $D$-term is undefined as was observed for the electron in QED \cite{Berends:1975ah,Milton:1976jr},
charged pions in chiral perturbation theory \cite{Kubis:1999db}, in computations of quantum gravitational 
corrections for charged spin-0 and spin-$\frac12$ particles 
\cite{Donoghue:2001qc} and proton \cite{Varma:2020crx,Mejia:2025oip}. The electron EMT was  recently further 
investigated in \cite{Metz:2021lqv,Freese:2022jlu}. Other systems with long-range forces were discussed 
in \cite{Panteleeva:2023aiz,Ji:2022exr,Czarnecki:2023yqd,Freese:2024rkr}. 
The goal of this work is to present a study of the EMT properties in the classical 
electron model constructed by Bia\l ynicki-Birula in \cite{Bialynicki-Birula:1983ace}. 

Since the discovery of the electron by Thomson in 1897 \cite{Thomson}, 
numerous efforts were made to construct classical models of extended, charged 
particles with contributions by Thomson, Abraham, Lorentz, Poincar\'e, Einstein, 
Wien, Planck, Sommerfeld, Langevin, Ehrenfest, Born, Pauli, von Laue and others
\cite{Bialynicki-Birula:1983ace}.
Even after quantum mechanics superseded classical physics in the 1920s, 
the formulation of consistent classical models of finite-size charged particles 
remained an intellectual challenge and continues attracting interest to this day
\cite{Bialynicki-Birula:1983ace,Dirac:1962iy,Schwinger:1983nt,Pearle,Boyer:1982ns,Bialynicki-Birula:1993shm,Rohrlich}.

In the model of Bia\l ynicki-Birula, the electron is made of a ``perfect, charged 
fluid and the electromagnetic field''~\cite{Bialynicki-Birula:1983ace}.
The model was introduced to help resolve a controversy in literature between Boyer 
and Rohrlich on how to consistently calculate the electromagnetic energy and momentum 
of a classical moving charged particle \cite{Boyer:1982ns,Rohrlich:1982nt}. 
In this work we will explore the model to study the EMT properties of the electron. 
Several features make this model an attractive choice for our purposes. 
(i) The model describes a finite-size electrically charged particle 
in a theoretically consistent and relativistic way. 
(ii) The cohesive forces which bind the system are included in the ``internal dynamics 
of the system''  \cite{Bialynicki-Birula:1983ace}. 
(iii) The simplicity of the model will allow us to compute the EMT properties
analytically and present lucid interpretations of the results.
In this model, the electron has a finite radius which in principle can be fixed 
to be smaller than the experimental upper bounds for size of the electron, but
in this work we will refrain from fixing parameters and focus on
analytical results.

It is expected that results from classical electrodynamics 
and QED should coincide in the long-distance limit \cite{Donoghue:2001qc}.
The goal of this work is to test this expectation.  
After introducing the model in Sec.~\ref{Sec-2:model-electron},
we will study the spatial distribution related to the energy density 
and the stress tensor distributions in Sec.~\ref{Sec-3:EMT-in-model},
and compute the form factors $A(t)$ and $D(t)$ in the model in
Sec.~\ref{Sec-3:FFs-A-D}, which will include a study of the limits 
four-momentum transfer $t\to0$ and $R\to0$ 
to restore a pointlike electron. As an interesting byproduct, we will 
explore the model in Sec.~\ref{Sec-4:Dreg} to shed some light on the 
concept of a ``regularized $D$-term'' proposed for the proton 
\cite{Varma:2020crx,Mejia:2025oip}.
In Sec.~\ref{Sec-5:conclusions} we present the conclusions.

\newpage
\section{The Classical Model of the Electron}
\label{Sec-2:model-electron}

In this section, we introduce the classical electron model \cite{Bialynicki-Birula:1983ace}. 
In this model, the electron is made of a perfect charged fluid and the electromagnetic field. 
The energy momentum tensor (EMT) of the system is given by \cite{Bialynicki-Birula:1983ace}
\be\label{Eq:EMT}
    T^{\mu \nu} 
    = (\rhofl + \pfl) u^\mu u^\nu 
    + \frac{1}{4\pi} {F^\mu}_{\lambda} F^{\lambda \nu} 
    - g^{\mu \nu} \;\frac{1}{16\pi} F_{\alpha\beta}F^{\alpha\beta}\; ,
\ee
where $\rhofl$ and $\pfl$ are the mass-energy density and pressure of
the fluid and $u^\mu$ is its four-velocity, while $F^{\mu\nu}$ is the 
electromagnetic field strength tensor. The flow of the fluid is 
adiabatic, and $\rhofl$ and $\pfl$ as functions of the scalar 
fluid density $n$ obey the relation
\begin{equation}
    \frac{d\rhofl}{dn} = \frac{\rhofl + \pfl}{n}. \label{Eq:adiabatic-flow}
\end{equation}
The equations of motion describing the perfect charged fluid coupled to the 
electromagnetic field are given by 
\ba 
    \mbox{(a)} \quad
    q\,n\,F^{\alpha\nu} u_{\alpha} 
        + (\partial^{\nu} -u^{\nu} u^{\mu} \partial_{\mu})\pfl  
    = (\rhofl + \pfl) u^{\mu} \partial_{\mu} u^{\nu} ,  
    \quad \quad
    \mbox{(b)} \quad   
    \partial_\mu F^{\mu \nu} 
    =&4\pi\,q\,n\,u^\nu \, \label{Eq:EOM-1}
\ea
where $q\,n\,u^\mu$ is the four-current density with $q$  
the electron charge.\footnote{\label{Footnote-1} We use Gauss units 
    where the Coulomb force between two static charges $q_i$ is 
    $F=\frac{q_1q_2}{r^2}$ and natural units with $c=\hbar=1$. 
    In this unit system, the fine structure constant $\alpha=q^2\simeq\frac1{137}$, 
    i.e.\ the electron charge is given by $q= -\sqrt{\alpha}$.}
In the rest frame of the system, the fluid is described by a static configuration
and $u^\mu=(1,\,0,\,0,\,0)$ such that the equations of motion in Eq.~(\ref{Eq:EOM-1}) 
become
\ba
    \mbox{(a)} \quad
    q\,n\,\vec{E} - \vec{\nabla} \pfl = 0 \, , \quad \quad
    \mbox{(b)} \quad
    \vec{\nabla}\cdot\vec{E} = 4\pi\,q\,n\, . \label{Eq:EOM-2}  
\ea
In order to overcome the repulsion within the fluid and form a stable 
and bound system, the following equation of state was imposed in the model 
where $\kappa$ is a positive constant \cite{Bialynicki-Birula:1983ace},
\be
    \pfl(n) = -\kappa\,n^{6/5}. \label{Eq:EOS} 
\ee
The negative sign in Eq.~(\ref{Eq:EOS}) indicates that this pressure is  
a Poincar\'e stress, i.e.\ provides the cohesive force which is responsible for 
binding the system 
\cite{Poincare}.
Based on the equation of state (\ref{Eq:EOS}), the adiabatic flow 
equation (\ref{Eq:adiabatic-flow}) yields 
\be
    \rhofl(r) = \mfl \, n(r) - 5\,\kappa\,n(r)^{6/5}
\ee
where $\mfl$ is an integration constant due to solving 
Eq.~(\ref{Eq:adiabatic-flow}) with the meaning of the mechanical mass of 
the system, i.e.\ the mass the perfect fluid would have if it was not 
subject to electromagnetic forces and the equation of state (\ref{Eq:EOS}).
The solution of the static equations of motion in Eq.~(\ref{Eq:EOM-2}) 
is then given by 
\ba
    \vec{E}(\vec{r}) &=& E(r)\,\vec{e}_r \, , \quad
    E(r) = \frac{q\,r}{(R^2+r^2)^{3/2}} \, , \label{Eq:sol-E} \\
    n(r) &=& \frac{3}{4\pi}\;\frac{R^2}{(R^2+r^2)^{5/2}} \, ,\quad
    R^2 = \frac{4\pi}{3}\;\biggl(\frac{q^2}{6\kappa}\biggr)^{\!5} \, ,
    \label{Eq:sol-n} 
\ea
where $r=|\vec{r}|$ and $\vec{e}_r = \vec{r}/r$. The parameter $R>0$ 
characterizes the size of the system and can be referred to as the 
``electron radius'' \cite{Bialynicki-Birula:1983ace}. Notice that 
the scalar density is normalized as $\int d^3r\,n(r) = 1$ and 
Eq.~(\ref{Eq:EOM-2}a) is satisfied only for the value of $R$ quoted 
in Eq.~(\ref{Eq:sol-n}) while Eq.~(\ref{Eq:EOM-2}b) holds for any $R$.  

\newpage

\section{The spatial EMT distributions in the electron model}
\label{Sec-3:EMT-in-model}

After the preparations in Sec.~\ref{Sec-2:model-electron}, we are now
in the position to study the EMT properties of the electron in the classical 
model. In the following, we evaluate the components of the EMT in 
Eq.~(\ref{Eq:EMT}) in the rest frame of the system.

\subsection{Energy density \boldmath $T_{00}(r)$}

Inserting in Eq.~(\ref{Eq:EMT}) the solutions (\ref{Eq:sol-E},~\ref{Eq:sol-n}) 
we obtain for the energy density the result 
\ba\label{Eq:T00}
	T_{00}(r) 
        &=&
            \rhofl(r) + \frac{1}{8\pi}\,E(r)^2
        =   \frac{1}{8\pi} \biggl[
            \frac{6\mfl R^2}{(r^2+R^2)^{5/2}} 
          + \frac{\alpha \,r^2}{(r^2+R^2)^3}
          - \frac{\lambda_p\, \alpha \, R^2}{(r^2+R^2)^3} \biggr]
\ea
where we used the fine structure constant $\alpha = q^2$ 
(see footnote~\ref{Footnote-1} on units), and eliminated $\kappa$ by means of 
Eq.~(\ref{Eq:sol-n}). The latter step renders the model expressions  elegant
but makes it difficult to tell apart the terms of electromagnetic origins 
and those due to Poincar\'e stress in Eq.~(\ref{Eq:EOS}). 
We therefore introduce the constant
\be
    \lambda_p = 1 \, \label{Eq:lambda}
\ee
to label the contribution due to the Poincar\'e stress. The advantage of introducing 
$\lambda_p$ becomes immediately apparent when we compute the mass of the system, 
\be\label{Eq:M}
	M   = \int d^3r\,T_{00}(r)
        = \mfl + \frac{3\pi\,\alpha}{32R} 
        - \lambda_p\;\frac{5\pi\,\alpha}{32R}\,,
\ee
where we see in detail how $M$ arises: the mechanical mass $\mfl$ is augmented 
by the electromagnetic contribution $\frac{3\pi\,\alpha}{32R}$ because the electric 
field of the charge configuration carries energy, and is diminished by 
$-\,\lambda_p\frac{5\pi\,\alpha}{32R}$ due to the Poincar\'e stress which provides
the binding effect. Inserting $\lambda_p=1$ we see that the system is ultimately bound
\be\label{Eq:Mfinal}
	M = \mfl + E_{\rm bind} < \mfl \, ,  \quad \quad 
    E_{\rm bind} = - \frac{\pi\,\alpha}{16R} < 0 \,.
\ee

So far the electron appears point-like in experiments. The determinations of 
experimental upper bounds on its size are in general model dependent, see for 
example \cite{Brodsky:1980zm,Dehmelt:1990zz,Kopp:1994qv,L3:2000bql,Peskin:2020fug}.
For instance, the upper bound for the electron radius derived from data obtained
from Large Electron-Positron (LEP) collider experiments is
$R_{\rm exp} < 3.1\times 10^{-19}\rm m$ at a 95\% confidence level
\cite{L3:2000bql}. If we would set the model parameter $R=10^{-20}\rm m$ within 
experimental limits, then the electron mass arises from significant cancellations,
\be\label{Eq:Mnumeric}
    M =28,234.715\,{\rm MeV} + 42,351.306\,{\rm MeV} - 70,585.510\,{\rm MeV}=0.511\,{\rm MeV},
\ee 
with the different contributions quoted in the order of their appearance in 
Eq.~(\ref{Eq:M}). Notice that the physical mass is nearly negligible compared 
to the other numbers in Eq.~(\ref{Eq:Mnumeric}), and to a good approximation 
$\mfl \simeq \frac{\pi\alpha}{16R}= |E_{\rm bind}|$.

A realistic description of the electron substructure, if there was any, 
would of course require quantum field theoretical compositeness models. 
The point is that one can choose $R$ small enough for the model to not 
contradict experimental results. For $r\lesssim R$ the results are model dependent, but in the long-distance limit $r\gg R$ we will be able to use this classical framework 
to obtain model-independent results and confirm QED predictions. In the following, we 
will not need to fix the parameters and will focus on analytical results.

\begin{figure}[t!]
  \begin{center}
    \includegraphics[width=0.25\textwidth]{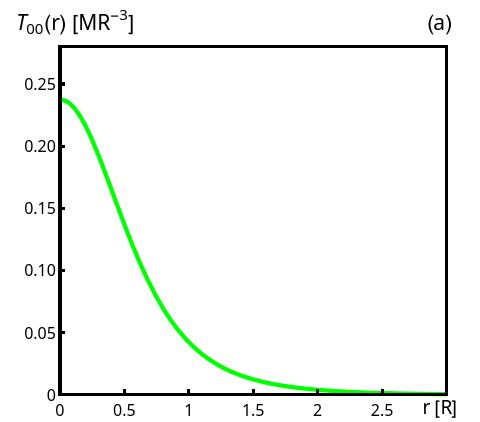} \ \
    \includegraphics[width=0.25\textwidth]{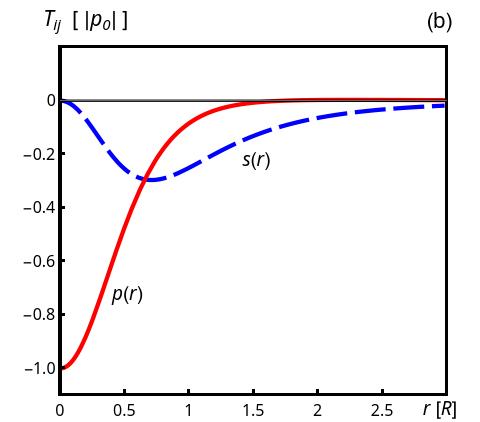} \ \ 
    \includegraphics[width=0.25\textwidth]{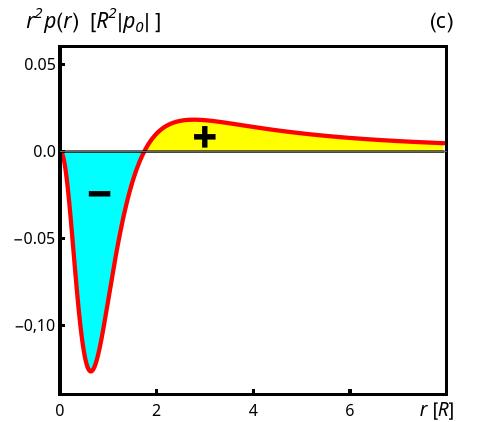}
  \end{center}
  \caption{\label{Fig-1:distributions}
    Electron model EMT distributions as functions of $r$ in units of electron radius $R$.
    (a) Energy distribution $T_{00}(r)$ in units of $M/R^3$.
    (b) The stress tensor distributions $s(r)$ and $p(r)$ in units of $|p_0|$.
    (c) The distribution of $r^2p(r)$ in units of $R^2|p_0|$ illustrating how the von Laue 
    condition in Eq.~(\ref{Eq:von-Laue}b) is satisfied. The sign patterns of the stress 
    tensor distributions $s(r)$ and $p(r)$ are opposite to what has been found in 
    hadronic models based on short-range interactions.}	
\end{figure}

In Fig.~\ref{Fig-1:distributions}a we show $T_{00}(r)$ in units of $M/R^3$ plotted 
as function of $r$ in units of~$R$. 
As can be seen in Fig.~\ref{Fig-1:distributions}a, the electron is a diffuse 
object in the model: 50$\,\%$ of its mass is contained within 
a radius of $r\le 1.3\,R$; and 80$\,\%$ within $r\le 2.5\,R$. 
Comparing $T_{00}(r)$ to the charge distribution $\rho_{el}(r)=q\,n(r)$,
we notice that
$T_{00}(r)/M-n(r)={\cal O}(\alpha)$. Therefore, $\rho_{el}(r)$ 
(plotted in units of $q/R^3$) and $T_{00}(r)$ (plotted in units 
of $M/R^3$) would not be distinguishable on the scale of 
Fig.~\ref{Fig-1:distributions}a due to the numerical smallness of $\alpha$. 
However, the long-distance behaviors of the two distributions differ with 
$T_{00}(r) \propto \frac1{r^4}$ while $\rho_{el}(r)\propto\frac1{r^5}$ for 
$r\gg R$. As a consequence the mean radii defined as 
\be
    \la r^k_E\ra=\frac{\int d^3r\,r^k T_{00}(r)}{M} \, , \quad
    \la r^k_{\rm ch}\ra=\frac{\int d^3r\,r^k \rho_{\rm ch}(r)}{q} \,   
    \label{Eq:radii-E}
\ee
diverge for the energy distribution for $k\ge 1$ and for the charge distribution for 
$k\ge 2$. Only the mean charge radius $\la r^1_{\rm ch}\ra=2R$ is finite.

\subsection{The \boldmath $T_{0k}$ components}
\label{Sec-3B:T0k}

The evaluation of the $T_{0k}$ components of the EMT in Eq.~(\ref{Eq:EMT}) 
for the solutions (\ref{Eq:sol-E},~\ref{Eq:sol-n}) yields $T_{0k}(\vec{r}) = 0$.
The classical angular momentum $J^i=\int d^3r\,\epsilon^{ijk}r^j T^{0k}$ of the system 
is therefore zero. Thus, in this toy model the ``electron'' appears to be spinless, 
although we shall find later an indication that the model is compatible with describing
a spin-$\frac12$ fermion rather than a spin zero boson (see Sec.~\ref{Sec-3:FFs-A-D}).
We remark that in principle one could try to explore projection techniques known from soliton 
physics \cite{Rajamaran} to assign spin quantum numbers to the classical solution. Note that
spin effects were of no importance for the purposes of Ref.~\cite{Bialynicki-Birula:1983ace}
and will also play no role in our study.

\subsection{The stress tensor \boldmath $T_{ij}$}

The static stress tensor $T^{ij}$ of a spin-0 or spin-$\frac12$ particle can be decomposed 
on general grounds into a traceless part and a trace described, respectively, in terms of 
the shear force $s(r)$ and pressure $p(r)$ distributions as follows
\be
        T^{ij}(\vec{r}) = 
            \biggl( e_{r}^{i}e_{r}^{j}-\frac{1}{3}\delta ^{ij}\biggr)s(r)
          + \delta ^{ij}\,p(r)\,.\label{Eq:Tij}
\ee
EMT conservation implies the following two relations \cite{Polyakov:2018zvc}
\be
    \mbox{(a)} \quad \frac23\,s^\prime(r)+\frac{2}{r}\,s(r)+p^\prime(r)=0\,, \quad \quad
    \mbox{(b)} \quad \int_0^\infty dr\,r^2 p(r)=0 \,. 
    \label{Eq:von-Laue}
\ee
The latter is known as the von Laue condition and is a necessary (but not sufficient) 
condition for stability \cite{Laue:1911lrk}. The computation of $s(r)$ and $p(r)$ in the model yields
\begin{alignat}{5}
        s(r)    &=  -\,\frac{1}{4\pi}\,E(r)^2
                &=& -\,\frac{\alpha}{4\,\pi}\;\frac{r^2}{(r^2+R^2)^3}\,,\label{Eq:s} \\
        p(r)    &=  \phantom{-\,}  \frac{1}{24\pi}\,E(r)^2 - \lambda_p\,\kappa n(r)^{6/5}
                &=& \phantom{-\,}  \frac{\alpha}{24\,\pi}\;\frac{r^2}{(r^2+R^2)^3} 
            -       \lambda_p\,\frac{\alpha}{8\,\pi}\;\frac{R^2}{(r^2+R^2)^3} \,. \label{Eq:p} 
\end{alignat}
It can be readily checked analytically that the model results for $s(r)$ and $p(r)$ in
Eqs.~(\ref{Eq:s},~\ref{Eq:p}) comply with the conditions in Eq.~(\ref{Eq:von-Laue}), which is
an important internal consistency check for the model. The model results for the shear and
pressure distributions are shown in Fig.~\ref{Fig-1:distributions}b with $s(r)$ and $p(r)$ 
plotted in units of the modulus of $p_0=p(0)$. 

Notice that $p_0=-\frac{\alpha}{8\pi R^4}$ is negative and therefore can be expressed as 
${p_0 = \frac{2E_{\rm bind_{ }}}{\pi^2R^3}}$ with the (negative) binding energy $E_{\rm bind}$ 
in Eq.~(\ref{Eq:Mfinal}). Recall that pressure has the dimension of force/area or energy/volume, 
and the characteristic energy for stress tensor properties is the (in our case) negative binding 
energy. This is interesting. Another interesting observation is that $s(r)$ and $p(r)$ inside 
electron exhibit opposite signs to what have been observed in the interiors of hadrons 
in hadronic models based on short-range strong forces only (see the reviews 
\cite{Polyakov:2018zvc,Burkert:2023wzr}). All patterns are the same as in hadronic models, 
except for the overall sign reversal. For instance, the shear force distribution is $s(r)>0$ 
for $0<r<\infty$ in hadronic models, but has the opposite sign in the electron model. Similarly, 
in hadronic models $p(r)>0$ is positive in the center of the particle and negative in the outer region, 
while in the electron model it is vice versa. 
In the electron model and others, the pressure exhibits a single node as is typical for ground state solutions 
and necessary to comply with the von Laue condition in Eq.~(\ref{Eq:von-Laue}b). 
The node of $p(r)$ is located at $r=\sqrt{3}R$ and difficult to see in Fig.~\ref{Fig-1:distributions}b. 
We illustrate this feature and the compliance with the von Laue condition with more clarity by plotting 
$r^2p(r)$ in Fig.~\ref{Fig-1:distributions}c.

At this point, we have no definite explanation for the reversed sign pattern of $s(r)$ and $p(r)$ in the 
electron model as compared to hadronic models. It is important to stress that the model describes the
electron as a perfectly stable bound state as proven in \cite{Bialynicki-Birula:1983ace}. However, the
cohesive forces are introduced in this model by imposing the equation of state in Eq.~(\ref{Eq:EOS}) 
in an ad-hoc way, as has been criticized (by the same author!) in Ref.~\cite{Bialynicki-Birula:1993shm}. 
A possible explanation could therefore be that this is too simplistic a model with ad-hoc Poincar\'e 
stresses. A perhaps more plausible explanation might be that this effect is due to the long-range nature of 
electric forces. EMT properties different from those in hadrons have also been observed also in other systems 
governed by long-range forces  \cite{Panteleeva:2023aiz,Ji:2022exr,Czarnecki:2023yqd,Freese:2024rkr}.

However, our goal is not to investigate the ``inner structure of the electron'' 
in the experimentally unexplored region ${r\lesssim 10^{-20}\,\rm m}$. Rather, our primary 
scope is to study the long-distance effects at distances $r\gg R = 10^{-20}\,\rm m$. From the 
results in Eqs.~(\ref{Eq:T00},~\ref{Eq:s},~\ref{Eq:p}) we read off the long-distance behavior 
of the EMT distributions
\be
    T_{00}(r)   =    \frac{\alpha}{ 8\pi}\;\frac{1}{r^4} + \dots\,,\quad \quad
    s(r)        = -\,\frac{\alpha}{ 4\pi}\;\frac{1}{r^4} + \dots\,,\quad \quad
    p(r)        =    \frac{\alpha}{24\pi}\;\frac{1}{r^4} + \dots\,
    \label{eq:long-distance}
\ee
where the dots indicate subleading terms decaying at large 
distances like $\frac1{r^5}$ or faster. 
The results in Eq.~(\ref{eq:long-distance}) agree with 
Ref.~\cite{Donoghue:2001qc,Varma:2020crx,Mejia:2025oip,Metz:2021lqv,Freese:2022jlu}
and offer a ``mathematical'' explanation of the reversed sign pattern as follows. 

For $r\gg R$, the model must (and does) reproduce the electric field of a pointlike charge with 
the electrostatic energy density $T_{00}(r) = \frac{1}{8\pi}\vec{E}^2$ and Maxwell stress tensor 
$T_{ij}(r) = -\,\frac1{4\pi}(E_iE_j-\frac12\delta_{ij}\vec{E}{ }^{\,2})$ which also yields
Eq.~(\ref{eq:long-distance}). Thus, the signs at asymptotically large distances are dictated by 
Maxwell's theory of electromagnetism and there must be no modification to the model in the far 
region. Modifications can occur only in the model-dependent near region, and since $p(r)$ must 
have a node somewhere (in the near region) to comply with the von Laue condition 
(\ref{Eq:von-Laue}b),  one is inevitably lead to the sign pattern of $p(r)$ in 
Fig.~\ref{Fig-1:distributions}. The sign pattern of $s(r)$ then follows from
Eq.~(\ref{Eq:von-Laue}a).
In the next section, we will discuss the consequences for the EMT form factors due 
to the long-distance behavior of the EMT distributions in Eq.~(\ref{eq:long-distance}).

\section{\boldmath The EMT form factors $A(t)$ and $D(t)$}
\label{Sec-3:FFs-A-D}

The matrix elements of the total symmetric EMT operator $\hat{T}^{\mu\nu}$ 
of a quantum particle (in the absence of parity violating interactions) 
define the EMT form factors as \cite{Kobzarev:1962wt,Ji:1996ek}
\ba
    \la p^\prime| \hat T^{\mu\nu} |p\rangle
    = \bar u(p^\prime,s^\prime)\biggl[
      A(t)\,\frac{P^\mu\gamma^\nu+P^\nu\gamma^\mu}{2}
    + B(t)\ \frac{i\,(P^\mu\sigma^{\nu\rho}+P^{\nu}\sigma^{\mu\rho})\Delta_\rho}{4M}
    + D(t)\,\frac{\Delta^\mu\Delta^\nu-g^{\mu\nu}\Delta^2}{4M}\biggr]u(p,s) \,,
    \label{Eq:EMT-FFs-def} 
\ea
where $P=\frac12(p+p^\prime)$ and $\Delta=p^\prime-p$. The spinor normalization is
$\bar u(p,s^\prime)u(p,s)=2M\,\delta_{s^{ }s^\prime}$ and the general constraints
$A(0)=1$ and $B(0)=0$ hold due to Lorentz invariance.
For particles whose size $R$ is much larger than their Compton wavelength $\lambda_c\gg 1/M$, 
i.e.\ $RM\gg 1$, the Fourier transform of the EMT matrix elements in the Breit frame where 
$p^\mu=(E,-\frac12\vec{\Delta})$ and $p^{\prime\,\mu}=(E,\frac12\vec{\Delta})$ gives rise 
to spatial distributions according to the interpretation \cite{Polyakov:2002yz}
\be\label{Eq:def-static-EMT}
    T^{\mu\nu}(\vec{r}) = \int\frac{d^3\Delta}{2E(2\pi)^3}\,
    \la p^\prime,s^\prime| \hat T^{\mu\nu} |p,s\ra\,
    e^{-i\vec{\Delta}\cdot\vec{r}}\,.
\ee
In classical models, the spatial distributions can be exactly computed, and form factors 
are obtained from inverting the Fourier transforms \cite{Varma:2020crx,Mejia:2025oip}. 
$D(t)$ can be obtained in two different ways: from $s(r)$ and $p(r)$, which we denote as
$D_s(t)$ and $D_p(t)$ and which must coincide. $A(t)$ is determined
in a unique way. The expressions are given by
\cite{Goeke:2007fp,Goeke:2007fq,Cebulla:2007ei}
\ba
&&  D_s(t) = 4M \int d^3r\,\frac{j_2(r\sqrt{-t})}{t}\,s(r) \, , \quad
    D_p(t) = 6M \int d^3r\,\frac{j_0(r\sqrt{-t})}{t}\,p(r) \, , 
    \label{Eq:DFF}\\
&&  A(t)   = \frac{1}{M} \int d^3r\,j_0(r\sqrt{-t})\,T_{00}(r) 
           - \frac{t}{4M^2}\Bigl(B(t)-D(t)\Bigr) \, ,
    \label{Eq:AFF}\ea
where $j_i(x)$ are spherical Bessel functions.

The Compton wavelength of the electron is much larger than its experimentally 
known upper bound for its size, and the condition $RM\gg 1$ is not satisfied. 
In such a situation, one could argue that one may assume $RM\gg 1$ during the
calculation before taking the limit to realistic values. However, with our focus remaining
on analytical results, we are not going to insert the numerical values anyway. 
Evaluating the expressions for $D(t)$ in the model, we obtain
\ba\label{Eq:DFF-1}
&& D_s(t)=\frac{\alpha\,\pi\,M}{4}\,e^{-R \sqrt{-t}} 
          \,\frac{1}{\sqrt{-t}}\, \nonumber\\
&& D_p(t)=\frac{\alpha\,\pi\,M}{4}\,e^{-R \sqrt{-t}} 
          \left[\frac{3 (1-\lambda_p)}{4 R t}
          +\frac{(3 \lambda_p +1)}{4\sqrt{-t}}\right]\,.
\ea
$D_s(t)$ arises solely from the electric contribution, while $D_p(t)$ arises from 
both electric forces and Poincar\'e stresses.
Recalling that $\lambda_p$ is unity (see Eq.~(\ref{Eq:lambda})), we obtain the same 
expression $D(t)=D_s(t)=D_p(t)$ from either method as expected, which constitutes 
another theoretical consistency test of the model. Expanding $D(t)$ around $t=0$ yields
\be\label{Eq:DFF-2}
D(t)= \frac{\alpha\,\pi}{4}\,\frac{M}{\sqrt{-t}} - \frac{\alpha\,\pi}{4}\,R\,M  + \dots 
\ee
The leading term in Eq.~(\ref{Eq:DFF-2}) is model independent and agrees with the QED result 
\cite{Freese:2022jlu}. The subleading terms are model dependent as can be seen from the 
appearance of the model parameter $R$. The dots in Eq.~(\ref{Eq:DFF-2}) indicate terms 
proportional to $\sqrt{-t}$ and higher powers thereof. In small-$t$ expansion of $D(t)$ 
in QED a term $\ln\sqrt{-t}$ appears at subleading order \cite{Freese:2022jlu}, 
which does not appear in the model. 

In order to compute $A(t)$ in Eq.~(\ref{Eq:AFF}) we need, in addition to $T_{00}(r)$ 
and $D(t)$ (which are known in the model), the form factor $B(t)=A(t)-2J(t)$ where 
$J(t)$ is the EMT form factor associated with total angular momentum (which is not known
in the model, see Sec.~\ref{Sec-3B:T0k}). Since $B(t)$ vanishes at $t=0$ and moreover 
enters in Eq.~(\ref{Eq:AFF}) with a prefactor of $t$, this contribution becomes relevant 
for $A(t)$ in Eq.~(\ref{Eq:AFF}) only beyond ${\cal O}(t)$ and can be safely neglected
if we study $A(t)$ in the small-$t$ region. The computation of $B(t)$ is beyond the 
capability of the present model.  

Evaluating the Fourier transform of $T_{00}(r)$ in Eq.~(\ref{Eq:AFF}) and combining with the
result for $D(t)$ from $s(r)$ in Eq.~(\ref{Eq:DFF-1}) (where no $\lambda_p$ appears) yields 
the following result
\be\label{Eq:AFF-1}
     A(t) = \alpha\,\pi
       \biggl[\frac{3-5\lambda_p}{32MR}-\frac{(3 + 5\lambda_p)\sqrt{-t}}{32M}\biggr]e^{-R\sqrt{-t}}
     + \frac{\mfl}{M}\, \Bigr[R\sqrt{-t}\,K_1(R\sqrt{-t}\,)\Bigl] \;+\; \frac{tB(t)}{4M^2}\,,
\ee
where $K_n(z)$ is the modified Bessel function of the second kind for $n=1$. Expanding this result around 
$t=0$ with $\sqrt{z}\,K_1(\sqrt{z})= 1+\frac{1}{4} z(\ln z + 2\gamma - 1 - 2 \ln 2) + {\cal O}(z^{3/2})$,
where $\gamma$ is the Euler-Mascheroni constant, yields 
\be\label{Eq:AFF-2}
     A(t) = 1 - \frac{3\,\pi\,\alpha}{16} \;\frac{\sqrt{-t}}{M} 
     - \frac{\mfl}{2M}\;R^2\,t\,\ln\Bigl(\tfrac12R\sqrt{-t}\Bigr) + \dots
\ee
where the dots indicate terms of ${\cal O}(t)$. We made use of Eq.~(\ref{Eq:M}) 
to simplify the coefficients of the first two terms. The model complies with the 
general constraint $A(0)=1$. The first non-trivial term in Eq.~(\ref{Eq:DFF-2})
proportional to $\sqrt{-t}$ corresponds exactly to the QED result \cite{Freese:2022jlu}.
We see that the model generates a subleading logarithmic term in $A(t)$
as in QED \cite{Freese:2022jlu} but with a different, model-dependent coefficient. 
Higher order terms are also model dependent. 

It is also instructive to consider the small $R$ limit. Expanding the model expressions
in Eqs.~(\ref{Eq:DFF-1},~\ref{Eq:AFF-1}) at fixed $t$ for small $R$ yields the results
\ba
    A(t) = 1 - \frac{3\,\pi\,\alpha}{16} \;\frac{\sqrt{-t}}{M}  
    \Bigl\{1 + {\cal O}(\epsilon\ln\epsilon)\Bigr\}\,, \quad \quad
    D(t)  =  \frac{\alpha\pi}{4} \;\frac{M}{\sqrt{-t}} 
    \Bigl\{1 + {\cal O}(\epsilon)\Bigr\}\,, \quad \quad
    \epsilon = R\,\sqrt{-t} \;.
    \label{Eq:FF-3}
\ea
For $A(t)$, we used Eq.~(\ref{Eq:M}) and $\mfl = {\cal O}(\frac{\alpha}{R})$ which
follows from the discussion below Eq.~(\ref{Eq:Mnumeric}).
In the limit $R\to0$ the $\frac1R$ singularity in the electron mass $M$ 
is compensated by the ``counter term'' $\mfl$, which reflects the renormalization of the 
electron mass due to short-distance (UV) effects in QED. The small-$R$ limit mimics the 
restoration of a point-like particle in the model. The results in Eq.~(\ref{Eq:FF-3}) 
demonstrate that we can derive model-independently the leading terms in 
Eqs.~(\ref{Eq:DFF-2},~\ref{Eq:AFF-2}), but not higher order terms.

Next we consider the limit of a point-like, free particle, which can be carried out by
taking in Eq.~(\ref{Eq:FF-3}) the limit $R\to0$ followed by taking the electromagnetic 
coupling constant $\alpha\to 0$. The result is trivial
\be
        A(t) = 1 \, , \quad \quad
        D(t) = 0 \, . \label{Eq:FF-4}
\ee
For point-like, free particles one expects constant form factors 
\cite{Hudson:2017xug,Hudson:2017oul}. Since $A(0)=1$ must hold
in any (free or interacting) theory, one must expect $A(t)=1$ for point-like, 
non-interacting particles of all types which we obtain. For $D(t)$ the situation 
is different because the result depends on particle type, e.g., one has $D(t)=1$ for 
a free spin-0 boson \cite{Hudson:2017xug}, and $D(t)=0$ for a free spin-$\frac12$ fermion 
\cite{Hudson:2017oul}. Our result in Eq.~(\ref{Eq:FF-3}) is compatible with the $D(t)$ of
a free spin-$\frac12$ fermion which indicates that this model really describes an 
electron, even though the explicit description of spin degrees of freedom 
(if possible in this framework) would require additional effort (see Sec.~\ref{Sec-3B:T0k}).  


For completeness, we note that for large $(-t)$, the model form factors  
in Eqs~(\ref{Eq:DFF-1},~\ref{Eq:AFF-1}) decay exponentially and $\lim_{t\to-\infty}A(t)/D(t)=-1$.
This differs from QED \cite{Freese:2022jlu} and is expected as this limit is beyond the reach 
of classical models. Notice that with increasing $(-t)$ QED is eventually superseded by the 
electroweak theory. For discussions of EMT properties in electroweak theory for Higgs and $Z$-bosons, 
we refer to Ref.~\cite{Beissner:2025nmg}.

\section{\boldmath Comments on proton $D$-term regularization of Refs.~\cite{Varma:2020crx,Mejia:2025oip}}
\label{Sec-4:Dreg}

The leading $1/\sqrt{t}$-term in Eqs.~(\ref{Eq:DFF-2},~\ref{Eq:FF-3}) is a universal 
QED effect which implies that $D$-terms of charged particles are divergent and undefined.
In Refs.~\cite{Varma:2020crx,Mejia:2025oip} it was proposed that for the proton it makes sense 
to introduce a ``regularization'' to define a finite, regularized value $D_{\rm reg}$ of the 
$D$-term for the following reason. The proton is an extended particle and the shape of its $D(t)$ 
is strongly dominated by the strong forces in the experimentally accessible $t$-region,
and QED effects become noticeable only for $(-t)\ll 10^{-4}\,\rm GeV^2$ which cannot
be accessed in experiments in the foreseeable future \cite{Varma:2020crx,Mejia:2025oip}. 
In contrast to the proton, in the case of the electron the concept of a regularized $D$-term is 
not physically justified. However, it is interesting to explore the exactly solvable electron model 
to give insights on how this regularization method works. 

The regularization method \cite{Mejia:2025oip} works as follows. 
Let $F(r)$ be a 3D distribution related to the stress tensor, and let
the integral $I_n = \int d^3 r^n F(r)$ be divergent due to the long-range em forces. 
We define a regularized integral by  
\be\label{Eq:reg-method}
    I_{n,\rm reg} = \int d^3r\,r^n \,F(r)_{\rm reg}= \int d^3r\,r^n \biggl[F(r) 
    + Z_F\Bigl\{\tfrac23\,n\,s(r)+(n+3)\,p(r)\Bigr\}\biggr] , \quad
    Z_F = 
    \lim\limits_{r\to\infty} \frac{r^4 F(r)}{ \tfrac12(n-1)\,\frac{\alpha}{4\pi}}
    \;,
\ee 
where the added term in the curly brackets originates from the identity 
$\frac23 s'(r)+\frac2rs(r)+p'(r)=0$ with the quoted expression following
after an integration by parts. The coefficient $Z_F$ in Eq.~(\ref{Eq:reg-method}) 
is defined in such a way that the integral $I_n$ becomes finite. Let us apply this
method first to the $D$-term.

Taking the limit $t\to0$ in Eq.~(\ref{Eq:DFF}) yields the expressions
$D_s = -\frac{4}{15}\,M \int d^3r\,r^2\,s(r)$ and $D_p = M \int d^3r\,r^2\,p(r)$
which are both divergent due to the long-distance behavior of the distributions 
in Eq.~(\ref{eq:long-distance}). Applying the regularization method in 
Eq.~(\ref{Eq:reg-method}) we obtain from $D_s$ and $D_p$ the unambiguous result
\be\label{eq:Dreg2}
    D_{\rm reg} = -\,\lambda_p \,\frac{\alpha\,\pi}{4}\, M R\,.
\ee

Several comments are in order. First, the regularization works and ``removes'' the 
divergence due to electromagnetic effects leaving behind a finite, and negative value 
for the regularized $D$-term. Second, the appearance of $\lambda_p$ signals that the
result is due to the Poincar\'e stresses. This is in line with results from models based 
on short-range forces where the $D$-terms are finite, negative, and their negative
signs can be traced back to the forces binding the system. Third, comparing the 
small-$t$ expansion of $D(t)$ in the model in Eq.~(\ref{Eq:DFF-2}) and the result 
in Eq.~(\ref{eq:Dreg2}) and recalling that $\lambda_p=1$, we see what the regularization 
method did: it removed the first, divergent term in Eq.~(\ref{Eq:DFF-2}) which is
of electromagnetic origin and retained the second, finite term which originates 
from the Poincar\'e stresses. 

As a second example we consider the mechanical radius which is defined in terms of the 
normal force per unit area $p_n(r)=\frac23s(r)+p(r)$ as follows
\be\label{Eq:r-mech}
    \la r^2_{\rm mech}\ra = \frac{\int d^3r\,r^2 p_n(r)}{\int d^3r\,p_n(r)}\;.
\ee
In models based on short range forces the distribution $p_n(r)$ is positive definite,
making the definition (\ref{Eq:r-mech}) a meaningful proxy for particle size. In the 
electron model, the $p_n(r)$ is negative and the numerator in Eq.~(\ref{Eq:r-mech}) is divergent
at large-$r$ due to the long-distance asymptotics in Eq.~(\ref{eq:long-distance}).
Remarkably, the regularization prescription fixes both issues since 
\be\label{Eq:r-mech-2}
    \bigl[p_n(r)\bigr]_{\rm reg} = \frac{\alpha}{2\pi}\;\frac{\lambda_p R^2}{(r^2+R^2)^3}\ge 0, \quad
    \la r^2_{\rm mech}\ra = \frac{\int d^3r\,r^2 [p_n(r)]_{\rm reg}}{\int d^3r\,[p_n(r)]_{\rm reg}}\
    = 3\lambda_p R^2\,.
\ee
Note that both the numerator and denominator in Eq.~(\ref{Eq:r-mech-2}) explicitly depend on 
$\lambda_p$. Thus, it is the Poincar\'e stress that gives the electron in this model a mechanical 
radius within the regularization procedure of Eq.~(\ref{Eq:reg-method}).

We stress that it is not our intention to define a regularized $D$-term for the electron. 
But for the proton it arguably makes sense to introduce a $D_{\rm reg}$ for practical purposes 
\cite{Varma:2020crx,Mejia:2025oip}, and our investigation shows how this regularization method works: 
it removes the electromagnetic contribution from $D$ and leaves behind a negative contribution from 
internal binding forces. 

\newpage

\section{Conclusions}
\label{Sec-5:conclusions}

We have carried out a study of the EMT properties of the electron in a lucid classical model 
by Bia\l ynicki-Birula \cite{Bialynicki-Birula:1983ace}. In this model, the electron is made 
of a charged ideal fluid, electric forces and imposed Poincar\'e stresses designed to overcome 
the electrostatic repulsion in the fluid and yield a stable bound state. The parameters in the 
model can in principle be fixed such that the system has the mass $M$ of the electron and the 
size $R=10^{-20}\rm m$ within experimental limits of the electron radius. However, the model
can be solved exactly and the parameters $M$ and $R$ can be left general in all steps. 
The description of spin degrees of freedom is beyond the scope of this model.

In classical frameworks, one can compute exactly the EMT 3D spatial distributions.
The obtained model results are very interesting. The positive definite energy distribution
$T_{00}(r)$ does not indicate anything unusual, but the distributions associated with the
stress tensor exhibit exactly the opposite sign patterns to what is found in hadronic models 
governed by short-range strong forces only. We have shown that from a mathematical point of view,
this is an inevitable consequence of the fact that at large distances $r\gg R$ the model must 
reproduce the results from (classical or quantized) Maxwell theory of electromagnetism, which 
is the case. More work may be needed to understand the underlying physical reasons. 
Ultimately, the reversed sign pattern is related to the masslessness of the photon and the 
long-range character of the electromagnetic interaction. 

In quantum frameworks, one may compute form factors and interpret them in terms
of 3D spatial densities in systems where the particle's size is large compared to the 
particle's Compton wavelength which requires $MR\gg1$. For the electron this condition 
is not met. However, in our analytically carried out calculation we may, for instance, assume 
that $MR$ is kept large throughout the calculation, and invert the Fourier transforms in the 
quantum interpretation to compute form factors from exactly known, classical 3D distributions. 
The results are insightful. We have shown that the model complies with the general constraint 
$A(0)=1$ for the form factor $A(t)$ and moreover reproduces the leading non-analytic terms of 
$A(t)$ and $D(t)$, which are proportional to respectively $\sqrt{-t}$ and $1/\sqrt{-t}$ with 
correct coefficients that coincide with QED \cite{Metz:2021lqv,Freese:2022jlu}. This supports
the validity of the model results obtained for the EMT form factors in the small-$(-t)$ region.
One remarkable result is that in the limit of a point-like, non-interacting particle
one obtains in the model $D(t)=0$ which corresponds to a free spin $\frac12$-fermion.

As an application of this lucid model, we presented the investigation of the concept of 
a regularized proton $D$-term \cite{Varma:2020crx}. This concept is motivated by the fact that 
for the proton the QED divergence in $D(t)\sim 1/\sqrt{-t}$ is noticeable only for 
$(-t)\ll 10^{-4}\rm GeV^2$ beyond the reach of experiments \cite{Mejia:2025oip}, and one will see 
$D(t)_{\rm prot}\approx D_{\rm prot,\rm reg}/(1+t/m^2_D)^{n_D}$ with some parameters $m_D$ and 
$n_D$ in the experimentally accessible region.
In the proton case, the regularization is designed to ``remove'' the electromagnetic contribution 
to $D(t)$ without affecting the strong interaction contributions responsible for the binding 
effects as can be inferred from a comparison to the neutron \cite{Mejia:2025oip}.

In the electron case, the concept of a regularized $D$-term is not justified. However, it is
interesting to use the model to see how this regularization works. In fact, in the model 
the stress tensor distributions defining the $D$-term receive contributions from electric forces and
Poincar\'e stresses, and the regularization removes only the former leaving behind a result which is 
(i) finite, (ii) negative, and (iii) due to the Poincar\'e stresses which bind the system. 
This is in line with results from hadronic models based on short range forces only where one also
finds a $D$-term which is (i) finite, (ii) negative, and (iii) the sign of the $D$-term is ultimately
due to the forces which bind the system \cite{Goeke:2007fp}. 

It would be interesting to investigate whether a classical model can be constructed which 
is also capable of reproducing subleading terms in the small-$t$ expansion of the EMT form factors
in QED. Another interesting question is whether there is a way to include spin degrees of freedom.
These topics will be left to future studies.

\ \\
\noindent{\bf Acknowledgments.} 
The authors wish to thank Adam Freese, Barbara Pasquini and Simone Rodini for 
valuable discussions. This work was supported by the National Science Foundation 
under the Award No.\ 2412625, and DOE under the umbrella of the Quark-Gluon
Tomography (QGT) Topical Collaboration with Award No. DE-SC0023646.

\end{document}